\documentclass[12pt]{article}
\usepackage{amsmath}
\usepackage{amssymb}
\newsymbol \blackbox 1004
\newcommand{\eh}{\hfill}\newlength{\sperr}

\newenvironment{proof}{{\settowidth{\sperr}{\bf\rm
Proof}%
\par\addvspace{0.3cm}\noindent\parbox[t]{1.3\sperr}
{\bf\rm P\eh r\eh o\eh o\eh f\eh }%
}}{\nopagebreak\mbox{}
$\blackbox$\par\addvspace{0.3cm}}

\def\a{\alpha}
\def\b{\beta}

\def\G{\Gamma}

\def\s{\sigma}

\def\O{\Omega}

\def\vp{\varphi}

\def\wt{\widetilde}
\def\ov{\overline}
\def\p{\partial}
\def\op{\overline{\partial}}
\def\BC{{\mathbb C}}
\def\BR{{\mathbb R}}

\newtheorem{Pa}{Paper}[section]
\newtheorem{Tm}[Pa]{{\bf Theorem}}

\newtheorem{Cy}[Pa]{{\bf Corollary}}
\newtheorem{Rk}[Pa]{{\bf Remark}}

\newtheorem{Ee}[Pa]{{\bf Example}}
\newtheorem{Dn}[Pa]{{\bf Definition}}
\newtheorem{Pn}[Pa]{{\bf Proposition}}
\newcommand{\CC}
{{\mathchoice {\setbox0=\hbox{$\displaystyle\rm
C$}\hbox{\hbox
to0pt{\kern0.4\wd0\vrule height0.9\ht0\hss}\box0}}
{\setbox0=\hbox{$\textstyle\rm C$}\hbox{\hbox
to0pt{\kern0.4\wd0\vrule height0.9\ht0\hss}\box0}}
{\setbox0=\hbox{$\scriptstyle\rm C$}\hbox{\hbox
to0pt{\kern0.4\wd0\vrule height0.9\ht0\hss}\box0}}
{\setbox0=\hbox{$\scriptscriptstyle\rm C$}\hbox{\hbox
to0pt{\kern0.4\wd0\vrule height0.9\ht0\hss}\box0}}}}

\title{Harmonic maps, B\"acklund-Darboux transformations and "line soliton" analogs}

\author{Alexander Sakhnovich}

\date{}
\begin{document}
\maketitle

{\bf Short title.} \vspace{5mm} Harmonic Maps and GBDT


E-mail address: al$_-$sakhnov@yahoo.com\\

\begin{abstract}
Harmonic maps from $\BR^2$ or  one-connected domain $\O \subset
\BR^2$ into $GL(m, \BC)$ and $U(m)$ are treated. The GBDT version
of the B\"acklund-Darboux transformation is applied to the case of
the harmonic maps. A new general formula on the GBDT
transformations of the Sym-Tafel immersions is derived. A class of
the harmonic maps similar in certain ways to line-solitons is
obtained explicitly and studied.

\end{abstract}

{\bf MSC2000:}  53C43; 35Q51

{\it Keywords:} Harmonic map; B\"acklund-Darboux transformation;
immersion; line soliton

\section{Introduction} \label{intro}
\setcounter{equation}{0}

Harmonic maps are actively studied in differential geometry,
mathematical physics and soliton theory, and the famous
B\"acklund-Darboux transformation  can be fruitfully  used for
this purpose. Since the original works of B\"acklund and Darboux
various interesting versions of the  B\"acklund-Darboux
transformation have been introduced (see, for instance,  \cite{C,
DoK, GeH, Gu,  Mar, MS, Mi, T0, UhT2, Uh, ZM} and references
therein). Some of these versions have been successfully applied to
the studies of harmonic maps.

In our paper we shall consider harmonic maps from $\BR^2$ or
one-connected domain $\O \subset \BR^2$ into $GL(m, \BC)$ and
$U(m)$. Here $GL(m, \BC)$ is the Lie group of $m \times m$
invertible matrices with the complex valued entries, and $U(m)$ is
its subgroup of unitary matrices. Correspondingly the map
$u(x,y)\,$ ($ (x,y) \in \BR^2$) is called harmonic if it satisfies
the Euler-Lagrange equation
\begin{equation} \label{0.1}
\op\big((\p u)u^{-1}\big)+\p\big((\op u)u^{-1}\big)=0 \quad
\Big(\p=\frac{1}{2}\big(\frac{\p}{\p x}-i \frac{\p}{\p y}\big), \,
\op=\frac{1}{2}\big(\frac{\p}{\p x}+i \frac{\p}{\p y}\big)\Big).
\end{equation}
We  apply a version of the B\"acklund-Darboux transformation (so
called GBDT), developed in \cite{SaA1}-\cite{SaA6} and some other
works of the author, to the case of the harmonic maps. A new
general formula on the GBDT transformations of the Sym-Tafel
immersions is derived.  A class of the harmonic maps similar in
certain ways to line-solitons is obtained explicitly and studied.

GBDT version of the B\"acklund-Darboux transformation for harmonic
maps is constructed in  Section \ref{GBDT}. In particular, GBDT
transformations of the Sym-Tafel immersions are given in
Proposition \ref{immers}. Explicit solutions are treated in
Section \ref{expl}.

\section{GBDT version of the B\"acklund-Darboux
transformation} \label{GBDT} \setcounter{equation}{0}

Suppose $u$ and its partial derivatives are continuously
differentiable and $u$ is a harmonic map into $GL(m, \BC)$. Then
Euler-Lagrange equation (\ref{0.1}) is equivalent \cite{Po} to the
compatibility condition
\begin{equation} \label{0.2}
\op G(x,y, \lambda) - \p F(x,y, \lambda)+[G(x,y, \lambda),F(x,y,
\lambda)]=0, \quad ([G,F]:=GF-FG)
\end{equation}
for system
\begin{equation} \label{0.3}
\p w(x,y, \lambda)=G(x,y, \lambda)w(x,y, \lambda), \quad \op
w(x,y, \lambda)=F(x,y, \lambda)w(x,y, \lambda),
\end{equation}
where
\begin{equation} \label{0.4}
G(x,y, \lambda)=-(\lambda-1)^{-1}q(x,y), \quad F(x,y,
\lambda)=-(\lambda+1)^{-1}Q(x,y),
\end{equation}
\begin{equation} \label{0.4'}
q(x,y)=(\p u(x,y))u(x,y)^{-1}, \quad Q(x,y)=-(\op
u(x,y))u(x,y)^{-1}.
\end{equation}
Without loss of generality assume $(0,0) \in \O$. To construct
GBDT fix an integer $n>0$ and five parameter matrices $A_1$,
$A_2$, $S(0,0)$, $\Pi_1(0,0)$ and $\Pi_2(0,0)$, where $A_1$,
$A_2$, $S$ are $n \times n$ matrices, and $\Pi_1$, $\Pi_2$ are $n
\times m$ matrices. We require also that $\pm 1 \not\in \s(A_k)$
($k=1,2$, $\s$ - spectrum) and the identity
\begin{equation} \label{0.5}
A_1S(0,0)-S(0,0)A_2=\Pi_1(0,0)\Pi_2(0,0)^*
\end{equation}
holds. Now introduce matrix functions $\Pi_1(x,t)$ and
$\Pi_2(x,t)$ by their  values at $(x,y)=(0,0)$ and linear
differential equations
\begin{equation} \label{0.6}
\p \Pi_1=(A_1-I_n)^{-1}\Pi_1q, \quad \op
\Pi_1=(A_1+I_n)^{-1}\Pi_1Q,
\end{equation}
\begin{equation} \label{0.7}
\p \Pi_2^*=-q\Pi_2^*(A_2-I_n)^{-1}, \quad \op
\Pi_2^*=-Q\Pi_2^*(A_2+I_n)^{-1},
\end{equation}
where $I_n$ is the $n \times n$ identity matrix. Similar to the
case of differentiation in real valued arguments in \cite{SaA3,
SaA6} the compatibility of both systems (\ref{0.6}) and
(\ref{0.7}) follows from the compatibility condition (\ref{0.2}).

Next, introduce $m \times m$ matrix function $S(x,y)$ via $S(0,0)$
and the partial derivatives
\begin{equation} \label{0.8}
\p S=-(A_1-I_n)^{-1}\Pi_1q\Pi_2^*(A_2-I_n)^{-1}, \quad \op
S=-(A_1+I_n)^{-1}\Pi_1 Q \Pi_2^*(A_2+I_n)^{-1}.
\end{equation}
From (\ref{0.2}) and (\ref{0.6})-(\ref{0.8}) it follows that $\op
\p S=\p \op S$, i.e., $S_{xy}=S_{yx}$, where $S_x=\frac{\p S}{\p x
}=\p S+ \op S$, $S_y=\frac{\p S}{\p y }=i(\p S- \op S)$. Thus the
entries of $S_x$ and $S_y$ form potential fields and so equations
(\ref{0.8}) are compatible.

According to (\ref{0.6})-(\ref{0.8}) we have $\p \big( A_1S-SA_2
\big) =\p \big( \Pi_1\Pi_2^* \big)$ and $\op \big( A_1S-SA_2 \big)
=\op \big( \Pi_1\Pi_2^* \big)$, which in view of (\ref{0.5})
implies the identity
\begin{equation} \label{0.9}
A_1S(x,y)-S(x,y)A_2=\Pi_1(x,y)\Pi_2(x,y)^*.
\end{equation}
Assume now that $S$ is invertible and consider well known in
system theory transfer matrix function represented in the Lev
Sakhnovich form \cite{SaL1}-\cite{SaL3}:
\begin{equation} \label{0.10}
w_A(x,y, \lambda)=I_m-\Pi_2(x,y)^*S(x,y)^{-1}(A_1-\lambda
I_n)^{-1}\Pi_1(x,y).
\end{equation}
The matrix function $w_A$ is invertible \cite{SaL1}:
\begin{equation} \label{0.10'}
w_A(x,y, \lambda)^{-1}=I_m+\Pi_2(x,y)^*(A_2-\lambda
I_n)^{-1}S(x,y)^{-1}\Pi_1(x,y).
\end{equation}
Formula (\ref{0.10'}) easily follows from the identity
(\ref{0.9}).

From Theorem 1 \cite{SaA3} (see also Appendix A \cite{SaA4}) it
follows that $w_A$ is a so called Darboux matrix function:
\begin{Pn} \label{PnGBDT} Suppose $m\times m$ matrix function $u$ and its partial
derivatives are continuously differentiable and relations
(\ref{0.1}), (\ref{0.4'}), and (\ref{0.5})-(\ref{0.8}) are valid.
Then in the points of invertibility of $S$ we have
\begin{equation} \label{0.11}
\p w_A(x,y, \lambda)= \wt G(x,y, \lambda)w_A(x,y, \lambda)-
w_A(x,y, \lambda)G(x,y, \lambda),
\end{equation}
\begin{equation} \label{0.12}
\op w_A(x,y, \lambda)= \wt F(x,y, \lambda)w_A(x,y, \lambda)-
w_A(x,y, \lambda)F(x,y, \lambda),
\end{equation}
where $w_A$ is given in  (\ref{0.10}), $G$ and $F$ are defined by
(\ref{0.4}),
\begin{equation} \label{0.13}
\wt G(x,y, \lambda)=-(\lambda-1)^{-1}\wt q(x,y), \quad \wt F(x,y,
\lambda)=-(\lambda+1)^{-1}\wt Q(x,y),
\end{equation}
\begin{equation} \label{0.14}
\wt q(x,y)=w_A(x,y, 1) q(x,y) w_A(x,y, 1)^{-1},
\end{equation}
\begin{equation} \label{0.15}
 \wt
Q(x,y)=w_A(x,y, -1) Q(x,y) w_A(x,y, -1)^{-1}.
\end{equation}
\end{Pn}
Suppose  $w$ is an $m \times m$ invertible matrix function
satisfying (\ref{0.3})-(\ref{0.4'}).  By (\ref{0.3}) and
(\ref{0.4}) the equalities
\begin{equation} \label{0.15!}
\p w(x,y,0)=q(x,y)w(x,y,0), \quad \op w(x,y,0)=-Q(x,y)w(x,y,0)
\end{equation}
hold. So in view of (\ref{0.4'}) one can normalize $w$ so that
\begin{equation} \label{0.15!!}
w(x,y,0)=u(x,y),
\end{equation}
where $u$ satisfies the Euler-Lagrange equation (\ref{0.1}).
Normalized in this way matrix function $w(x,y, \lambda)$ is called
an extended (and corresponding to $u$) solution of the
Euler-Lagrange equation or extended frame.
\begin{Tm} \label{TmGBDT} Suppose $u$ satisfies the Euler-Lagrange
equation (\ref{0.1}), the conditions of Proposition \ref{PnGBDT}
are fulfilled, and $w$ is an extended corresponding to $u$
solution. Then the matrix function
\begin{equation} \label{0.15!!!}
\wt u(x,y):=w_A(x,y,0)u(x,y)
\end{equation}
also satisfies the Euler-Lagrange equation. Moreover the matrix
function
\begin{equation} \label{0.17}
 \wt w(x,y, \lambda):= w_A(x,y, \lambda) w(x,y, \lambda).
\end{equation}
is an extended solution such that $ \wt w(x,y,0)=\wt u(x,y)$.
\end{Tm}
\begin{proof}.
According to formulas (\ref{0.3}) and (\ref{0.17})  and to
Proposition \ref{PnGBDT} we have
\begin{equation} \label{0.16}
\p \wt w(x,y, \lambda)= \wt G(x,y, \lambda) \wt w(x,y, \lambda),
\quad \op \wt w(x,y, \lambda)= \wt F(x,y, \lambda) \wt w(x,y,
\lambda).
\end{equation}
From (\ref{0.16}) it follows that the compatibility condition
\begin{equation} \label{0.18}
\op  \wt G(x,y, \lambda) - \p  \wt F(x,y, \lambda)+[ \wt G(x,y,
\lambda), \wt F(x,y, \lambda)]=0
\end{equation}
is fulfilled. Condition (\ref{0.18}) can be rewritten in the form
\[
( \lambda+1) \op \wt q -( \lambda-1)   \p  \wt Q+[ \wt Q, \wt
q]=0,
\]
i.e., the coefficients of the polynomial in $\lambda$ on the
left-hand side of the last equation turn to zero. In other words
we have
\begin{equation} \label{0.19}
 \op \wt q(x,y) =   \p  \wt Q(x,y)=-\frac{1}{2}[ \wt Q(x,y), \wt q(x,y)].
\end{equation}
From (\ref{0.15!!})-(\ref{0.17}) it follows that $ \wt
w(x,y,0)=\wt u(x,y)$. Therefore taking  into account (\ref{0.13})
and (\ref{0.16}) we get
\begin{equation} \label{0.20}
 \p \wt u(x,y) =   \wt q(x,y)  \wt u(x,y), \quad  \op \wt u(x,y) = -  \wt Q(x,y)  \wt u(x,y).
\end{equation}
Hence in view of (\ref{0.19}) the matrix function $\wt u$
satisfies (\ref{0.1}).  Now we see that by (\ref{0.13}),
(\ref{0.16}), and (\ref{0.20}) $\wt w$ is an extended solution
corresponding to $\wt u$.
\end{proof}
Theorem \ref{TmGBDT} presents a GBDT method to construct harmonic
maps $\wt u$ into $GL(m, \BC)$ and corresponding extended
solutions. According to (\ref{0.10}) the choice of the eigenvalues
of $A_1$ defines simple and multiple poles of the Darboux matrix
$w_A$, and in this way our result is related to the interesting
papers \cite{An,  IZ, WARD} on the pole data for the soliton
solutions. Notice that we don't require parameter matrix $A_1$ to
be similar to diagonal (it may have an arbitrary Jordan
structure).

Consider the case $uu^*=u^*u=I_m$, i.e., $u$ is a harmonic map
into $U(m)$. Then we have $u_x u^*+uu_x^*=u_y u^*+uu_y^*=0$.
Therefore from (\ref{0.4'}) it follows that
\begin{equation} \label{0.21}
q^*=\frac{1}{2}\Big(uu_x^*+iuu_y^* \Big)=-
\frac{1}{2}\Big(u_xu^*+iu_yu^* \Big)=Q.
\end{equation}
Put now $A_1=-A_2^*=A$ (i.e., assume $A_1=-A_2^*$). Then, taking
into account (\ref{0.6}), (\ref{0.7}), and (\ref{0.21}) we can
assume $\Pi_2=\Pi_1$ and denote $\Pi_1$ by $\Pi$. Further in this
section we assume
\begin{equation} \label{0.22}
A_1=-A_2^*=A, \quad \Pi_1(x,t)=\Pi_2(x,t)=\Pi (x,t), \quad
{\mathrm{and}} \quad S(0,0)=S(0,0)^*.
\end{equation}
Hence, using (\ref{0.8}) and (\ref{0.21}) we obtain $(\p S)^*=\op
S$, and so $S_x=S_x^*$, $S_y=S_y^*$. The last equality in
(\ref{0.22}) now implies
\begin{equation} \label{0.23}
S(x,y)=S(x,y)^*.
\end{equation}
\begin{Cy} \label{CyU} Suppose the equality $u^*=u^{-1}$, the
conditions of Theorem \ref{TmGBDT}, and assumptions (\ref{0.22})
are true. Then we have
\begin{equation} \label{0.24}
\wt u(x,y)^*=\wt u(x,y)^{-1}, \quad \wt q(x,y)^*=\wt Q(x,y).
\end{equation}
\end{Cy}
\begin{proof}.
Identity (\ref{0.9}) and definition (\ref{0.10}) now take the form
\begin{equation} \label{0.24'}
AS(x,y)+S(x,y)A^*=\Pi(x,y)\Pi(x,y)^*,
\end{equation}
\begin{equation} \label{0.25}
 w_A(x,y,
\lambda)=I_m-\Pi(x,y)^*S(x,y)^{-1}(A-\lambda I_n)^{-1}\Pi(x,y).
\end{equation}
Formula (\ref{0.10'}) takes the form
\begin{equation} \label{0.26}
w_A(x,y, \lambda)^{-1}=I_m-\Pi(x,y)^*(A^*+\lambda
I_n)^{-1}S(x,y)^{-1}\Pi(x,y).
\end{equation}
In view of (\ref{0.23}), (\ref{0.25}),  and  (\ref{0.26}) we
derive
\begin{equation} \label{0.27}
w_A(x,y, \lambda)^{-1}=w_A(x,y,- \ov{ \lambda})^*
\end{equation}
In particular, we have $w_A(x,y, 0)^{-1}=w_A(x,y, 0)^*$ and the
first equality in (\ref{0.24}) follows from the definition of $\wt
u$. Moreover,  by (\ref{0.27}) we have $w_A(x,y, -1)^{-1}=w_A(x,y,
1)^*$ and so  the second equality in (\ref{0.24}) follows from
(\ref{0.14}), (\ref{0.15}), and (\ref{0.21}).
\end{proof}
\begin{Rk} \label{RkUni}
The uniton solutions have been introduced in the seminal paper
\cite{Uh}. Put
\begin{equation} \label{Rk1}
A_1=aI_n, \quad A_2=bI_n, \quad \pi=(a-b)^{-1}\Pi_2^*S^{-1}\Pi_1.
\end{equation}
We shall  derive some relations for $\pi$ to compare with the
limiting uniton case. By (\ref{0.9}) and (\ref{Rk1}) we easily get
\begin{equation} \label{Rk2}
\pi^2=(a-b)^{-2}\Big(\Pi_2^*S^{-1}A_1\Pi_1-\Pi_2^*A_2S^{-1}\Pi_1
\Big)=\pi,
\end{equation}
i.e., $\pi(x,y)$ is a projector. From  (\ref{0.7}) and (\ref{0.8})
it follows that
\begin{equation} \label{Rk3}
\p\big(  \Pi_2^*S^{-1} \big)=-\wt q  \Pi_2^*S^{-1}(A_2-I_n)^{-1}.
\end{equation}
Hence, taking into account (\ref{0.6}) we have
\begin{equation} \label{Rk4}
\p\big(  \Pi_2^*S^{-1}\Pi_1 \big)=-\wt q
\Pi_2^*S^{-1}(A_2-I_n)^{-1}\Pi_1+\Pi_2^*S^{-1}(A_1-I_n)^{-1}\Pi_1q.
\end{equation}
In view of (\ref{0.10}),  (\ref{0.10'}), (\ref{0.14}), and
(\ref{Rk1}) we obtain
\begin{equation} \label{Rk5}
\wt q=\Big(I_m- \frac{a-b}{a-1}\pi \Big)q\Big(I_m+
\frac{a-b}{b-1}\pi \Big).
\end{equation}
Using (\ref{Rk1}),  (\ref{Rk2}),  and (\ref{Rk5}) rewrite
(\ref{Rk4}) as
\begin{equation} \label{Rk6}
\p  \pi=-(b-1)^{-1} \frac{a-1}{b-1}\Big(I_m- \frac{a-b}{a-1}\pi
\Big)q \pi +(a-1)^{-1}\pi q.
\end{equation}
When $ a=\ov a, \quad b=-a$, we can assume $\Pi_1=\Pi_2$, $S=S^*$,
i.e., $\pi=\pi^*$ is an orthogonal projector and
\begin{equation} \label{Rk8}
\p  \pi= \frac{1-a}{(a+1)^2}\Big(I_m- \frac{2a}{a-1}\pi \Big)q \pi
+(a-1)^{-1}\pi q.
\end{equation}
The so called singular B\"acklund transformation that transforms
unitons into unitons deals with the case $a\to -1$. In this case
we have $(a+1)^{-1} \to \infty$, and so (\ref{Rk8}) implies
$(I_m-\pi)q\pi=0$ (see \cite{Uh} and formula (5.130) \cite{Gu}).
Similar transformations are possible for the equation
\[
\op\big(  \Pi_2^*S^{-1}\Pi_1 \big)=-\wt Q
\Pi_2^*S^{-1}(A_2+I_n)^{-1}\Pi_1+\Pi_2^*S^{-1}(A_1+I_n)^{-1}\Pi_1
Q.
\]
\end{Rk}

 It is of interest to consider harmonic maps into $SU(m)$.
For that purpose introduce the notion of minimal realization. Any
rational $m \times m$ matrix function $\varphi$ that tends to $D$
at infinity can be presented in the form
\begin{equation} \label{0.28}
\vp(\lambda)=D-C(A- \lambda I_r)^{-1}B,
\end{equation}
where $D$ is an $m \times m$ matrix, $C$ is an $m \times r$
matrix, $B$ is an $r \times m$ matrix,  and $A$  is an $r \times
r$ matrix, $r \geq 0$, and the case $r=0$ corresponds to $\vp
\equiv D$. This type representation is called a {\it realization}
in system theory.
\begin{Dn} \label{Dnreal}
Realization  (\ref{0.28}) is called minimal if the order $r$ of
$A$ is the minimal possible. This order is called the McMillan
degree of $\vp$.
\end{Dn}
Realization remains minimal under small perturbations of $A$, $B$,
$C$.
\begin{Tm} \label{Tmreal}
Suppose the conditions of Corollary \ref{CyU} are fulfilled,
$\s(A) \cap \s(-A^*)= \emptyset$,  and realization (\ref{0.25}) is
minimal for some $(x_0,y_0) \in \O$. Then for each $(x,y) \in \O$
we have
\begin{equation} \label{0.29}
\det \, w_A(x,y, \lambda)=\prod_{k=1}^n\frac{\lambda +\ov
{a_k}}{\lambda - {a_k}},
\end{equation}
where $\{a_k\}$ are the eigenvalues of $A$ taken with their
algebraic multiplicity.
\end{Tm}
\begin{proof}. It is well known that as $\s(A) \cap \s(-A^*)= \emptyset$ and
(\ref{0.25}) is a minimal realization of $w_A(x_0,y_0, \lambda)$,
so $\det\, w_A(x_0,y_0, \lambda)$ has poles in all the eigenvalues
of $A$ and only there. Moreover, if $\det \, D \not= 0$, then the
McMillan degrees of $\vp$ and $\vp^{-1}$ coincide. Thus
(\ref{0.26}) is a minimal realization of $w_A(x_0,y_0,
\lambda)^{-1}$. Hence $\det\, w_A(x_0,y_0, \lambda)^{-1}$ has
poles in all the eigenvalues  of $-A^*$ and only there. Therefore,
if all the eigenvalues of $A$ are simple we get
\begin{equation} \label{0.30}
\det \, w_A(x_0,y_0, \lambda)=\prod_{k=1}^n\frac{\lambda +\ov
{a_k}}{\lambda - {a_k}}.
\end{equation}
If $A$ has multiple eigenvalues we prove (\ref{0.30}) by small
perturbations of $A$ (and corresponding perturbations of $S$ so
that the identity (\ref{0.24'}) preserves). Finally, notice that
$\displaystyle{\det \, w_A(x,y, \lambda)=\prod_{k=1}^r
\frac{\lambda - {\hat a_k}}{\lambda - {a_k}}}$ for each $(x,y)$
with possible arbitrariness in the choice of the set of
eigenvalues $a_k$ of $A$ and $ {\hat a_k}$ of $-A^*$. Taking into
account that $w_A$ is also continuous we derive (\ref{0.29}) from
(\ref{0.30}).
\end{proof}
The next Corollary is immediate.
\begin{Cy} \label{CySU} Suppose the conditions of Theorem \ref{Tmreal}
are fulfilled,  $\det\, (iA) \in \BR$, and $u \in SU(m)$. Then we
have $\wt u \in SU(m)$.
\end{Cy}

In the framework of his theory of "soliton surfaces" A. Sym
associates with integrable nonlinear systems  the corresponding
$\lambda$-families of immersions
\begin{equation} \label{0.31}
R(x,y,\lambda)=w(x,y, \lambda)^{-1}\frac{\p}{\p\lambda}w(x,y,
\lambda),
\end{equation}
where $w$ are extended solutions \cite{Sy}. On the other hand, our
version of the Darboux matrix, that can be used in numerous
important cases, including harmonic maps treated in this section,
admits representation (\ref{0.10}), where dependence on $\lambda$
is restricted to the resolvent $(A_1-\lambda I_n)^{-1}$. Thus
$w_A$ is easily differentiated in $\lambda$. The next proposition
expresses immersions (\ref{0.31}) generated by GBDT in terms of
$\Pi_1$, $\Pi_2$, and $S$.
\begin{Pn} \label{immers} Suppose extended solution $\wt w$ is given by equality
(\ref{0.17}), where Darboux matrix $w_A$ admits representation
(\ref{0.10}) and identity (\ref{0.9}) holds. Then immersion $\wt
R:=\wt w^{-1}\frac{\p}{\p\lambda}\wt w$ admits representation
\begin{equation} \label{0.32}
\wt R=R-w^{-1} \Pi_2^*(A_2-\lambda I_n)^{-1}S^{-1}(A_1-\lambda
I_n)^{-1}\Pi_1 w.
\end{equation}
\end{Pn}
\begin{proof}. From (\ref{0.17}) it easily follows that
\begin{equation} \label{0.33}
\wt R=R+w^{-1}w_A^{-1}\Big(\frac{\p}{\p\lambda} w_A\Big) w.
\end{equation}
By (\ref{0.10}) we have
\begin{equation} \label{0.34}
\frac{\p}{\p\lambda} w_A(x,y,
\lambda)=-\Pi_2(x,y)^*S(x,y)^{-1}(A_1-\lambda I_n)^{-2}\Pi_1(x,y).
\end{equation}
Taking into account (\ref{0.9}), (\ref{0.10'}), and (\ref{0.34})
one gets
\[
w_A^{-1}\frac{\p}{\p\lambda} w_A=\frac{\p}{\p\lambda}
w_A-\Pi_2^*(A_2-\lambda
I_n)^{-1}S^{-1}\Pi_1\Pi_2^*S^{-1}(A_1-\lambda I_n)^{-2}\Pi_1
\]\[
=\frac{\p}{\p\lambda} w_A-\Pi_2^*(A_2-\lambda
I_n)^{-1}S^{-1}\Big((A_1-\lambda I_n)S-S(A_2-\lambda
I_n)\Big)S^{-1}(A_1-\lambda I_n)^{-2}\Pi_1
\]
\begin{equation} \label{0.35}
=- \Pi_2^*(A_2-\lambda I_n)^{-1}S^{-1}(A_1-\lambda I_n)^{-1}\Pi_1
.
\end{equation}
Substitute now (\ref{0.35}) into  (\ref{0.33}) to get
(\ref{0.32}).
\end{proof}
Proposition \ref{immers} can be used to construct conformal CMC
immersions.
\section{Explicit solutions} \label{expl} \setcounter{equation}{0}
For the simple seed solutions (see, for instance, solutions given
by  formulas (\ref{1.1}) and (\ref{1.31})) equations (\ref{0.6})
and (\ref{0.24'}) can be usually solved explicitly, so that
equalities (\ref{0.15!!!}) and (\ref{0.25}) provide us with the
explicit expressions for harmonic maps. In this section we
consider several examples in greater detail. First similar to
\cite{Gu} we put $m=2$ and  take  the most simple seed solution of
(\ref{0.1}):
\begin{equation} \label{1.1}
\displaystyle{ u(x,y)= e^{(\tau\ov z-\ov \tau z)j}\in U(2), \quad
z=x+iy, \quad \tau \not=0, \quad j=\left[
\begin{array}{lr}
1 & 0 \\ 0 & -1
\end{array}
\right]. }
\end{equation}
Transformations of this seed solution are treated in Examples
\ref{Ee1}-\ref{Ee3}. Example \ref{Ee4} deals with the arbitrary
$m$ and seed solution of the form (\ref{1.31}). It is of interest
in Examples \ref{Ee1}, \ref{Ee3} that asymptotics of the
constructed maps differs in one particular direction and
non-diagonal entries tend to zero, except, possibly, in this
direction. (See relations (\ref{1.8}), (\ref{1.10}), (\ref{1.27}),
and (\ref{1.30}).)

For $u$ of the form (\ref{1.1}) by (\ref{0.4'}) we obtain
\begin{equation} \label{1.2}
q(x,y)=- \ov \tau j, \quad Q(x,y)=-  \tau j.
\end{equation}
Partition the matrix function $\Pi$ into columns
\begin{equation} \label{1.3}
\Pi(x,y)=\left[
\begin{array}{lr}
\Phi_1(x,y) & \Phi_2(x,y)
\end{array}
\right], \quad \Pi(0,0)=\left[
\begin{array}{lr}
f_1 & f_2
\end{array}
\right].
\end{equation}
Then according to (\ref{0.6}), (\ref{0.22}), and (\ref{1.2}) we
have
\begin{equation} \label{1.4}
\Phi_1(x,y)=\exp\Big(-\tau\ov z(A+I_n)^{-1}- \ov \tau
z(A-I_n)^{-1}\Big)f_1,
\end{equation}
\begin{equation} \label{1.5}
\Phi_2(x,y)=\exp\Big(\tau\ov z(A+I_n)^{-1}+ \ov \tau
z(A-I_n)^{-1}\Big)f_2.
\end{equation}

Consider now the case $n=1$, $A=a$ ($a\not=\pm 1$, $a \not=- \ov
a$), $f_1,f_2 \not=0$. Recall that the  one-soliton solution
treated in \cite{Gu} (formulas (5.101), (5.102)) had the following
behavior on the lines $z=x+iy= \mu t$ ($\mu \in \BC$, $\mu
\not=0$, $-\infty <t< \infty$):
\begin{equation} \label{1.5'}
\wt u =g u, \quad g(x,y)=c_{\pm}e^{|kt|} \big(I_2+o(1)\big) \quad
\mathrm{for} \quad t\to \pm \infty, \quad c_{\pm}\in \BR.
\end{equation}
Our next example proves quite different.

\begin{Ee} \label{Ee1} The GBDT transformation $\wt u$ of the seed
solution is given by formula (\ref{0.15!!!}):
\[\wt
u(x,y)=w_A(x,y,0)u(x,y),
\]
where $w_A$ is defined in (\ref{0.25}). Let us study $\wt u$ on
the lines $z= \mu t$. For $\Pi$ in the right hand side of
(\ref{0.25}), by (\ref{1.3})-(\ref{1.5}) we get
\begin{equation} \label{1.6}
\Pi(x,y)=\left[
\begin{array}{lr}
e^{-bt}f_1 & e^{bt}f_2
\end{array}
\right], \quad b:=(a-1)^{-1}\ov\tau\mu+(a+1)^{-1}\tau\ov\mu .
\end{equation}
Hence from (\ref{0.24'}) and (\ref{1.6}) it follows that
\begin{equation} \label{1.7}
S(x,y)=(a+\ov a)^{-1}\Pi(x,y)\Pi(x,y)^*=(a+\ov a)^{-1}\Big(
e^{-(b+\ov b)t}|f_1|^2+ e^{(b+\ov b)t}|f_2|^2\Big).
\end{equation}
Without loss of generality we can assume $b+\ov b \geq 0$. In view
of (\ref{0.25}), (\ref{1.6}), and (\ref{1.7}) for $b+\ov b>0$ we
have
\begin{equation} \label{1.8}
\lim_{t \to \infty}w_A(x,y,0)=\left[
\begin{array}{lr}
1 & 0 \\ 0 & -\ov a/a
\end{array}
\right],  \quad \lim_{t \to - \infty}w_A(x,y,0)=\left[
\begin{array}{lr}
 -\ov a/a & 0 \\ 0 & 1
\end{array}
\right].
\end{equation}
The asymptotics differs on the line $\G_0=\{z: \,z=\mu t\}$, where
$b+\ov b=0$, i.e.,
\begin{equation} \label{1.9}
\mu=\frac{i \tau c}{(\ov a -1)(a+1)}, \quad c= \ov c\not=0.
\end{equation}
Namely, on $\G_0$ we have
\begin{equation} \label{1.10}
w_A(x,y,0)=\frac{1}{a}\left[
\begin{array}{lr}
\a & \b(t) \\ \ov{\b(t)} &- \ov \a
\end{array}
\right],   \quad \a= \frac{a|f_2|^2- \ov a
|f_1|^2}{|f_1|^2+|f_2|^2 },
\end{equation}
\begin{equation} \label{1.11}
\b(t)=-\frac{(a+ \ov a)\ov {f_1} f_2}{|f_1|^2+|f_2|^2 }e^{2bt}
\quad \quad (\ov b=-b).
\end{equation}
\end{Ee}

Similar to the "line solitons" studied, in particular, for KP
(see, for instance, \cite{AbSe, Sat, VA}) the asymptotics of our
solution on $\G_0$ essentially differs from asymptotics along
other lines. Notice also that only on $\G_0$ the non-diagonal
entries of $\wt u$ do not decay to zero.

\begin{Ee} \label{Ee2}
This example deals with the case $n=2$, $A={\mathrm{diag}}\{a_1,\,
a_2\}$, where diag means diagonal matrix,  $a_k\not=\pm 1$, and
$\s(A) \bigcap \s(-A^*)= \emptyset$. By (\ref{1.3})-(\ref{1.5}) on
a line $z=\mu t$ we have
\begin{equation} \label{1.12}
\Pi(x,y)=\left[
\begin{array}{lr}
e^{-Bt}f_1 & e^{Bt}f_2
\end{array}
\right], \quad B={\mathrm{diag}}\{b_1,\, b_2\},
\end{equation}
\[
b_k:=(a_k-1)^{-1}\ov\tau\mu+(a_k+1)^{-1}\tau\ov\mu  \quad (k=1,2).
\]
 Supposing $b_1+\ov {b_1} \not=0$ we  choose the sign of $\mu$
 so that $b_1+\ov {b_1} >0$. Assume that $b_2+\ov
{b_2} >0$ too (the case $b_2+\ov {b_2} <0$ can be treated
similar).  Let  the entries $f_{12}$ and $f_{22}$ of $f_2$ be
nonzero and let $a_1\not=a_2$. Then in view of (\ref{0.24'}) and
(\ref{1.12}) we easily get
\begin{equation} \label{1.13}
\Pi(x,y)=\left[
\begin{array}{lr}
0 & e^{b_1 t}f_{12} \\ 0 & e^{b_2t}f_{22}
\end{array}
\right]+o(1), \quad t \to \infty,
\end{equation}
\begin{equation} \label{1.14}
S(x,y)^{-1}=\frac{1}{\det \, S(x,y)}\left(\left[
\begin{array}{lr}
\frac{|f_{22}|^2}{a_2+\ov{a_2}}e^{(b_2+\ov{b_2})t} &
-\frac{f_{12}\ov{f_{22}}}{a_1+\ov{a_2}}e^{(b_1+\ov{b_2})t}
\\- \frac{f_{22}\ov{f_{12}}}{a_2+\ov{a_1}}e^{(b_2+\ov{b_1})t} &
\frac{|f_{12}|^2}{a_1+\ov{a_1}}e^{(b_1+\ov{b_1})t}
\end{array}
\right]+o(1)\right),
\end{equation}
\begin{equation} \label{1.15}
\det \,
S(x,y)=(1+o(1))\frac{|f_{12}f_{22}|^2\big(|a_{1}|^2+|a_{2}|^2-a_1\ov{a_2}-a_2\ov{a_1}
\big)
}{(a_1+\ov{a_1})(a_2+\ov{a_2})|a_1+\ov{a_2}|^2}e^{(b_1+\ov{b_1}+b_2+\ov{b_2})t}.
\end{equation}
After some calculations the asymptotics of
$w_A(x,y,0)=I_2-\Pi^*S^{-1}A^{-1}\Pi$  follows from
(\ref{1.13})-(\ref{1.15}):
\begin{equation} \label{1.16}
\lim_{t \to \infty}w_A(x,y,0)=\left[
\begin{array}{lr}
1 & 0 \\ 0 & \displaystyle{\frac{\ov {a_1}\, \ov { a_2}}{a_1 \,
a_2}}
\end{array}
\right].
\end{equation}
\end{Ee}
If in the last example, where $A$ is diagonal, we put $a_1=a_2=a$
and $\det \Pi(0,0)\not= 0$, we easily get a trivial answer
$w_A(x,y,0)=-(\ov a / a)I_2$. The case, where $A$ is a Jordan box,
is far more interesting.
\begin{Ee} \label{Ee3} Suppose now that
\begin{equation} \label{1.17}
n=2, \quad A=\left[
\begin{array}{lr}
a & 1 \\ 0 & a
\end{array}
\right], \quad a\not=\pm 1, \quad a \not=- \ov a.
\end{equation}
It follows that
\begin{equation} \label{1.18}
\tau \ov \mu (A+I_2)^{-1}+\ov \tau  \mu (A-I_2)^{-1}=b I_2+c R,
\quad R=\left[
\begin{array}{lr}
0 & 1 \\ 0 & 0
\end{array}
\right],
\end{equation}
where $b$ is given by the second relation in (\ref{1.6}) and
\begin{equation} \label{1.19}
c=-((a-1)^{-2}\ov\tau\mu+(a+1)^{-2}\tau\ov\mu).
\end{equation}
According to (\ref{1.3})-(\ref{1.5}) and (\ref{1.18}) on a line
$z=x+iy=\mu t$ we have
\begin{equation} \label{1.20}
\Pi(x,y)=[e^{-bt}(I_2-c t R)f_1 \quad e^{bt}(I_2+c t R)f_2].
\end{equation}
In view of (\ref{1.17}) identity (\ref{0.24'}) takes the form
\begin{equation} \label{1.21}
(a+\ov a)S+\left[
\begin{array}{lr}
s_{21}+s_{12} & s_{22} \\ s_{22} & 0
\end{array}
\right]=\Pi\Pi^*,
\end{equation}
where $s_{kj}$ are the entries of $S$. From (\ref{1.20}) and
(\ref{1.21}) it is immediate that
\begin{equation} \label{1.22}
s_{22}(x,y)=(a+\ov a)^{-1}\Big( e^{-(b+\ov
b)t}|f_{21}|^2+e^{(b+\ov b)t}|f_{22}|^2 \Big),
\end{equation}
\[
s_{12}(x,y)=\ov{s_{21}(x,y)}=(a+\ov a)^{-1}
\]
\begin{equation} \label{1.23}
\times\Big( e^{-(b+\ov b)t}(f_{11}-ctf_{21})\ov{f_{21}}+e^{(b+\ov
b)t} (f_{12}+ctf_{22})\ov{f_{22}}-s_{22}(x,y)\Big),
\end{equation}
\[
s_{11}(x,y)=(a+\ov a)^{-1}
\]
\begin{equation} \label{1.24}
\times\Big( e^{-(b+\ov b)t}|f_{11}-ctf_{21}|^2+e^{(b+\ov
b)t}|f_{12}+ctf_{22}|^2-s_{12}(x,y)-s_{21}(x,y) \Big).
\end{equation}
Similar to  Example \ref{Ee1} assume $b+\ov b \geq 0$. Let
$f_{21}\not= 0$ and $f_{22}\not= 0$ for simplicity. First we shall
treat the lines, where $b+\ov b > 0$. Taking into account
(\ref{1.22})-(\ref{1.24}) by  standard calculations we derive
\begin{equation} \label{1.25}
\det \, S(x,y)=\big( 1+o(1)\big)(a+\ov a)^{-4}e^{2(b+\ov
b)t}|f_{22}|^4.
\end{equation}
One can easily see that
\begin{equation} \label{1.26}
\Pi^*S^{-1}A^{-1}\Pi=a^{-2}(\det \, S)^{-1}\Pi^*\left[
\begin{array}{lr}
s_{22} &- s_{12} \\ -s_{21} & s_{11}
\end{array}
\right]\left[
\begin{array}{lr}
a & -1 \\ 0 & a
\end{array}
\right] \Pi.
\end{equation}
From (\ref{1.22})-(\ref{1.26}) it follows that
\[
\Pi^*S^{-1}A^{-1}\Pi \to \left[
\begin{array}{lr}
0 & 0 \\ 0 & a^{-2}(a^2-\ov a^2)
\end{array}
\right] \quad {\mathrm{when}} \quad t \to \infty, \quad i.e.,
\]
\begin{equation} \label{1.27}
\lim_{t \to \infty}w_A(x,y,0)=\left[
\begin{array}{lr}
1 & 0 \\ 0 & (a^{-1}\ov a)^2
\end{array}
\right] \quad {\mathrm{for}} \quad b+ \ov b>0.
\end{equation}
As in Example \ref{Ee1}, the asymptotics of $w_A(x,y,0)$ differs
on the line $\G_0$, where $b+ \ov b=0$. For this line by
(\ref{1.22})-(\ref{1.24}) we obtain
\begin{equation} \label{1.28}
\det \, S=4(a+\ov a)^{-2}|cf_{21}f_{22}|^2t^2+k_1t+k_0.
\end{equation}
Taking also into account (\ref{1.26}) we have
\begin{equation} \label{1.29}
\Pi^*S^{-1}A^{-1}\Pi=4a^{-1}(a+\ov a)^{-1}|cf_{21}f_{22}|^2(\det
\, S)^{-1}\left[
\begin{array}{lr}
t^2+O(t) & O(t) \\ O(t) & t^2+O(t)
\end{array}
\right].
\end{equation}
Formulas (\ref{1.28}) and (\ref{1.29}) imply
\begin{equation} \label{1.30}
\lim_{t \to \infty}w_A(x,y,0)=-(\ov a/a)I_2  \quad {\mathrm{for}}
\quad b+ \ov b=0.
\end{equation}
Compare equalities (\ref{1.27}) and (\ref{1.30}). We would like to
mention here that  the extended solution $\wt w=w_Au \,$ has the
pole of order two at $\lambda=a$:
\[ \wt w(t, \lambda)= (a-\lambda)^{-2}\]
\[
\times (\det \, S(t))^{-1}\Pi(t)^* \left[
\begin{array}{c} s_{22}(t)  \\ -s_{21}(t)
\end{array}
\right]\left[
\begin{array}{lr}
e^{-bt}f_{21} & e^{bt}f_{22}
\end{array}
\right]+O\big((a-\lambda)^{-1}\big).
\]
\end{Ee}
Our next example deals with an arbitrary $m$ and a seed solution
somewhat more general than the one given in (\ref{1.1}). Namely we
put
\begin{equation} \label{1.31}
\displaystyle{ u(x,y)= e^{\ov z D- zD^*}\in U(m), \quad
D={\mathrm{diag}}\{ d_1, \, d_2, \ldots , d_m \}.}
\end{equation}
It follows that
\begin{equation} \label{1.32}
q(x,y)=-D^*, \quad Q(x,y)=-D.
\end{equation}
\begin{Ee}  \label{Ee4}
As in Example \ref{Ee1}, suppose $n=1$, $A=a$ ($a\not=\pm 1$, $a
\not=- \ov a$), $f_k \not=0$. Here $1 \leq k \leq m$. Then,  on a
line $z=\mu t$ taking into account (\ref{0.6}) and  (\ref{0.24'})
we obtain
\begin{equation} \label{1.33}
\Pi(x,y)=\left[
\begin{array}{lcr}
e^{-b_1 t}f_1 & e^{-b_2 t}f_2 & \ldots
\end{array}
\right], \quad b_k:=(a-1)^{-1}\ov{ d_k} \mu+(a+1)^{-1}d_k \ov\mu,
\end{equation}
\begin{equation} \label{1.34}
S(x,y)=(a+\ov a)^{-1}\sum_{k=1}^{m}|f_k|^2e^{-(b_k+\ov{b_k}) t}
\end{equation}
In the generic situation there exists only one natural number
$k_+$ and only one natural number $k_-$ such that
\begin{equation} \label{1.35}
\Re b_{k_+}= \min_{1 \leq k \leq m} \Re b_{k}, \quad \Re b_{k_-}=
\max_{1 \leq k \leq m} \Re b_{k}.
\end{equation}
Then by (\ref{1.33}) and  (\ref{1.35}) we have
\[
\lim_{t \to \infty} w_A(x,y,0)={\mathrm{diag}}\{ 1,  \ldots , 1,
-\ov a/a, 1, \ldots \}, \, {\mathrm{where}} \, -\ov a/a \,
{\mathrm{is}}\, {\mathrm{the}}\, k_+{\mathrm{th}} \,
{\mathrm{entry}},
\]
\[
\lim_{t \to - \infty} w_A(x,y,0)={\mathrm{diag}}\{ 1,  \ldots , 1,
-\ov a/a, 1, \ldots \}, \, {\mathrm{where}} \, -\ov a/a \,
{\mathrm{is}}\, {\mathrm{the}}\, k_-{\mathrm{th}} \,
{\mathrm{entry}}.
\]

\end{Ee}


\begin{thebibliography}{AGKS}
\bibitem{VA}
J. Villarroel, M.J. Ablowitz, On the initial value problem for the
KPII equation with data that do not decay along a line,
Nonlinearity 17:5 (2004) 1834-1866.
\end{thebibliography}
\end{document}